\documentclass[conference]{IEEEtran}
\usepackage[utf8]{inputenc}
\usepackage[T1]{fontenc}
\usepackage{comment}
\usepackage{cite}
\usepackage{graphicx}
\usepackage{booktabs}
\usepackage{multirow}
\usepackage{amsmath,amssymb,mathtools,bm}
\usepackage{amsthm}
\usepackage{bbm}

\usepackage{siunitx}
\usepackage{url}
\usepackage{xcolor}
\usepackage{caption}
\usepackage{subcaption}
\usepackage{algorithm}
\usepackage{algorithmic}
\usepackage{booktabs,multirow}
\usepackage[group-separator={,},group-minimum-digits=4]{siunitx}
\usepackage{subcaption}
\newtheoremstyle{boldthm}
  {\topsep}   
  {\topsep}   
  {\itshape}  
  {}          
  {\bfseries} 
  {.}         
  { }         
  {}          

\theoremstyle{boldthm}
\newtheorem{theorem}{Theorem}
\newtheorem{corollary}{Corollary}
\usepackage{tikz}
\usetikzlibrary{arrows.meta,positioning,calc}
\tikzset{
  block/.style={draw, rounded corners, minimum height=8mm, minimum width=20mm,
                inner sep=3pt, align=center, font=\footnotesize},
  line/.style={-Latex, thick},
  lab/.style={font=\scriptsize}
}

\DeclareMathOperator*{\argmax}{arg\,max}

\title{Semantic Communication with Hopfield Memories\vspace{-3mm}}
\vspace{-6mm}\author{
\IEEEauthorblockN{Karim Nasreddine\IEEEauthorrefmark{1}, Christo Kurisummoottil Thomas\IEEEauthorrefmark{2}, and Walid Saad\IEEEauthorrefmark{3}}
\IEEEauthorblockA{\IEEEauthorrefmark{1}Department of Electrical and Computer Engineering, American University of Beirut, Beirut, Lebanon}
\IEEEauthorblockA{\IEEEauthorrefmark{2}Department of Electrical and Computer Engineering, Worcester Polytechnic Institute, Worcester, MA, USA}
\IEEEauthorblockA{\IEEEauthorrefmark{3}Department of Electrical and Computer Engineering, Virginia Tech, Alexandria, VA, USA}
Email: kwn01@aub.edu, cthomas2@wpi.edu, walids@vt.edu\vspace{-5mm}
}
\usepackage[final]{microtype}
\begin{document}
\maketitle

\begin{abstract}

Traditional joint source-channel coding employs static learned semantic representations that cannot dynamically adapt to evolving source distributions. Shared semantic memories between transmitter and receiver can potentially enable bandwidth savings by reusing previously transmitted concepts as context to reconstruct data, but require effective mechanisms to determine when current content is similar enough to stored patterns. However, existing hard quantization approaches based on variational autoencoders are limited by frequent memory updates even under small changes in data dynamics, which leads to inefficient usage of bandwidth.
 To address this challenge, in this paper, a memory-augmented semantic communication 
framework is proposed where both transmitter and receiver maintain a shared memory of 
semantic concepts using modern Hopfield networks (MHNs). The proposed framework 
employs soft attention-based retrieval that smoothly adjusts stored semantic prototype weights 
as data evolves that enables stable matching decisions under gradual data dynamics. 
A joint optimization of encoder, decoder, and memory retrieval 
 mechanism is performed with the objective of maximizing a reasoning capacity metric that quantifies 
semantic efficiency as the product of memory reuse rate and compression ratio. 
Theoretical analysis establishes the fundamental rate-distortion-reuse tradeoff 
and proves that soft retrieval reduces unnecessary transmissions compared to 
hard quantization under bounded semantic drift. Extensive simulations over 
diverse video scenarios demonstrate that the proposed MHN-based approach 
achieves substantial bit reductions around $14\%$ on average and up to $70\%$ in scenarios 
with gradual content changes compared to baseline. Moreover,  the simulation results show that the proposed solution achieves a superior performance under transmission errors where baseline hard quantization 
methods fail abruptly.
\end{abstract}

\vspace{-2mm}\section{Introduction}\vspace{-1mm}
Traditional source coding techniques \cite{cover1991elements}, even those based on artificial intelligence (AI) \cite{farsad2018deep} typically transmit data representations tightly coupled to pixel-level fidelity for images, videos and word-level fidelity for text, thereby requiring substantial bandwidth even when the receiver is interested in only high-level semantic content. \emph{Semantic communication (SC)} \cite{chaccour2024less,lan2021semantic,strinati2021sixg,kountouris2021semantics} offers a fundamentally different paradigm by faithfully representing the meaning of the data by encoding the semantics into low-dimensional latent representations that preserve task-relevant information. A learned decoder at the receiver then reconstructs the original content from these compact semantic features~\cite{DeepSC,VQDeepSC}. In SC, the transmitter and receiver can share a semantic knowledge base which allows them to reconstruct the content using its local knowledge. This approach leverages advanced AI capabilities at the receiver to minimize over-the-air payload and achieve bandwidth efficiency beyond traditional joint source-channel coding (JSCC). However, such contextual knowledge-based SC faces  practical 
challenges such as (a) \emph{scalability}, as the number of stored semantic concepts 
grows with content diversity, (b) need for 
\emph{efficient knowledge retrieval and update mechanisms} that can adapt to 
dynamic changes in data while balancing bandwidth efficiency and 
reconstruction quality, 
and (c) guarantees of  \emph{semantic consistency} between transmitter and receiver 
knowledge bases under transmission and memory errors.  

\vspace{-2mm}\subsection{Related works}\label{sec:related}
\vspace{-2mm}
A number of recents works \cite{XieMemDeepSC2023, ChristoTWCArxiv2022,ren2023knowledge} explored  context-based and memory-augmented 
SC~\cite{XieMemDeepSC2023, ChristoTWCArxiv2022,ren2023knowledge}. The work in \cite{XieMemDeepSC2023} introduced the concept of semantic 
memory for communication systems, but focused primarily on transformer-based 
architectures without analyzing the fundamental tradeoffs between 
memory size, retrieval mechanisms, and communication cost. In \cite{ChristoTWCArxiv2022}, the authors proposed emergent semantic languages 
through signaling games, establishing shared knowledge bases between 
communicating agents. The work in \cite{hello2024semantic} proposed a knowledge base-assisted SC 
framework where graph neural networks (GNNs) are used to embed the stored relations among semantic concepts. Recent works like~\cite{fu2023vqsc} and \cite{miao2024vqdeepvsc} leverage vector quantized variational autoencoders 
(VQ-VAE)~\cite{oord2017neural} to learn discrete semantic codebooks for 
compact indexing and transmission, demonstrating significant compression 
gains over traditional deep JSCC. While these approaches demonstrate the potential 
of shared context in enhancing bandwidth efficiency, they face critical 
limitations. For instance, 
GNN inference for 
large-scale knowledge graphs incurs prohibitive latency,
while the need for building comprehensive semantic knowledge graphs requires extensive 
domain-specific ontologies and manual curation, limiting generalizability 
to dynamic or novel semantic concepts. Moreover, existing SC methods such as those in \cite{XieMemDeepSC2023, ChristoTWCArxiv2022,ren2023knowledge} do not account for knowledge base mismatches between transmitter and receiver, necessitating efficient memory retrieval and storage mechanisms.  


\vspace{-2mm}\subsection{Contributions}\vspace{-1.5mm}
The main contribution of this paper is to address the aforementioned challenges by proposing a novel memory-augmented 
SC framework that leverages modern Hopfield networks (MHNs) \cite{Ramsauer2021} as 
shared associative memories.  Associative memories use similarity-based retrieval rather than exact matching, providing robust recall under gradual data changes and transmission errors without domain-specific ontologies. Their soft attention mechanism scales efficiently as the number of semantic concepts increases.
 Although Hopfield networks have 
been successfully applied to perception and reasoning tasks, their potential 
for wireless communication systems, particularly for managing the semantic knowledge bases under bandwidth constraints and channel impairments remains unexplored. In summary, our key contributions are:
\begin{itemize}
    \item We develop a novel SC framework that integrates a shared associative memory using MHNs into the encoder–decoder system. 
    \item We introduce a new reasoning capacity metric (inspired from \cite{chaccour2024less}) that quantifies semantic communication efficiency as ``useful queries per bit.'' This metric connects memory reuse to an intelligence-per-bit measure and allows evaluation of how effectively the system converts transmitted bits into correct semantic retrievals.
    \item We provide a novel theoretical analysis characterizing the rate–distortion–reuse tradeoff. We derive conditions under which the Hopfield associative recall mechanism outperforms VQ's hard quantization, particularly near decision boundaries where small semantic drifts would cause VQ codebooks to proliferate. 
    \item Extensive 
    simulations conducted over diverse data characteristics demonstrate that MHN-driven SC achieves 
    approximately $14\%$ reduction in average bits per frame compared to VQ at 
    matched reconstruction quality (PSNR within $\pm$0.2~dB). Under dynamic scenarios, MHNs achieves $70\%$ bit savings 
    ($68.7$~bits/frame vs. $228.5$~bits/frame) by maintaining sustained reuse through 
    gradual content changes, while baseline methods trigger frequent unnecessary refreshes. 
\end{itemize}

\vspace{-1.5mm}\section{System Model}
\label{sec:system}
We consider a point-to-point SC system consisting of a 
transmitter (Tx) and a receiver (Rx) connected over a wireless channel. The 
Tx aims to communicate a video sequence $\{\bm{x}_t\}_{t=1}^T$ to the Rx, 
where $t$ indexes the frame number and each frame 
$\bm{x}_t \in \mathbb{R}^{H \times W \times 3}$ represents an RGB image of 
height $H$ and width $W$. The system objective is to minimize the average 
bits per frame while ensuring that the Rx's reconstruction $\hat{\bm{x}}_t$ 
satisfies a target distortion constraint $\mathbb{E}[D(\bm{x}_t, \hat{\bm{x}}_t)] \leq D_0$, 
where $D(\cdot, \cdot)$ is a perceptual distortion metric (e.g., mean squared 
error or structural similarity). 
\vspace{-2mm}\subsection{Encoder-decoder architecture}
A learned encoder $f_\theta : \mathbb{R}^{H \times W \times 3} \to \mathbb{S}^{d-1}$ 
maps each frame to a $d$-dimensional unit-norm latent vector 
$\bm{z}_t = f_\theta(\bm{x}_t)$, where $\mathbb{S}^{d-1} = \{\bm{z} \in \mathbb{R}^d : \|\bm{z}\|_2 = 1\}$ 
is the $(d-1)$-dimensional unit sphere and $\theta$ are the encoder 
parameters. A decoder $g_\phi : \mathbb{S}^{d-1} \to \mathbb{R}^{H \times W \times 3}$ 
reconstructs frames as $\hat{\bm{x}}_t = g_\phi(\hat{\bm{z}}_t)$ from the 
received latent representation $\hat{\bm{z}}_t$, where $\phi$ are the decoder 
parameters. The spherical latent space enables similarity-based retrieval via 
cosine similarity $\langle \bm{z}_i, \bm{z}_j \rangle$.

The \emph{shared associative memory} at time $t$ is a set of $M_t$ prototype 
vectors $
\mathcal{M}_{t} = \{\bm{\mu}_1,\ldots,\bm{\mu}_{M_t}\}, \bm{\mu}_j \in \mathbb{S}^{d-1},
$
where each prototype $\bm{\mu}_j$ represents a learned semantic concept, i.e., a 
recurring pattern in the latent space corresponding to perceptually similar 
content such as ``blue sky,'' ``person walking,'' or ``static background.'' 
Both Tx and Rx maintain identical copies of $\mathcal{M}_t$, which can be synchronized 
through the emergent language protocol discussed in \cite{ChristoTWCArxiv2022}. The memory grows dynamically as new semantic 
concepts are encountered, up to a maximum capacity $M_{\max}$. Prototypes can be learned through prior offline training on 
representative video data or emerge online through the adaptive protocol in ~\cite{ChristoTWCArxiv2022}.

\vspace{-3mm}\subsection{Transmission decision}\vspace{-1mm}

Given the current latent $\bm{z}_t$, the Tx queries the memory 
$\mathcal{M}_{t-1}$ by computing cosine similarities to all stored prototypes:
$
s_j = \langle \bm{z}_t, \bm{\mu}_j \rangle, \quad j \in [M_{t-1}],
$
and identifies the best-matching prototype $j^* = \argmax_{j \in [M_{t-1}]} s_j$ 
with maximum similarity $s_{\max} = s_{j^*}$. Using a similarity threshold 
$\tau \in (0,1)$, the Tx makes a binary reuse decision:
\vspace{-2mm}\begin{equation}
H_t = 
\begin{cases}
1, & \text{if } s_{\max} \geq \tau \quad \text{(hit: reuse stored prototype)}; \\
0, & \text{if } s_{\max} < \tau \quad \text{(miss: transmit new latent)}.
\end{cases}
\label{eq:hit_indicator}
\end{equation}
The transmission decision can be formalized as a policy 
$\pi : \mathbb{S}^{d-1} \times \mathcal{M}_{t-1} \to \{0,1\} \times [M_{t-1}]$ 
that maps the query-memory pair $(\bm{z}_t, \mathcal{M}_{t-1})$ to a hit/miss 
indicator $H_t$ and, if $H_t = 1$, the index $j^* \in [M_{t-1}]$ to transmit. 
On a miss ($H_t = 0$), the policy transmits the new latent $\bm{z}_t$ and 
may append it to memory as $\mathcal{M}_t = \mathcal{M}_{t-1} \cup \{\bm{z}_t\}$ 
if capacity permits ($M_{t-1} < M_{\max}$); otherwise, the memory remains 
unchanged ($\mathcal{M}_t = \mathcal{M}_{t-1}$).

The Rx reconstructs the latent as:
\vspace{-2mm}\begin{equation}
\hat{\bm{z}}_t = 
\begin{cases}
\bm{\mu}_{j^*}, & \text{if } H_t = 1 \quad \text{(retrieve from memory)}; \\
Q(\bm{z}_t) + \bm{\epsilon}_t, & \text{if } H_t = 0 \quad \text{(decode transmitted latent)},
\end{cases}
\label{eq:decoder_input}
\end{equation}
where $Q(\cdot) : \mathbb{S}^{d-1} \to \mathbb{S}^{d-1}$ is a quantizer 
(e.g., uniform scalar quantization with 8 bits per dimension followed by 
normalization), $\bm{\epsilon}_t \in \mathbb{R}^d$ is an optional residual 
for quality enhancement (typically $\bm{\epsilon}_t = \bm{0}$ in our 
experiments), and $\bm{\mu}_{j^*} \in \mathcal{M}_{t-1}$ is the retrieved 
prototype. The Rx then reconstructs the frame as 
$\hat{\bm{x}}_t = g_\phi(\hat{\bm{z}}_t)$.
The threshold $\tau$ controls the rate-distortion-reuse tradeoff: 
higher $\tau$ yields fewer hits (more new transmissions, higher rate, 
potentially lower distortion), while lower $\tau$ increases reuse (more 
index transmissions, lower rate) but accepts less similar matches that may 
require residual bits $\bm{\epsilon}_t$ to maintain reconstruction quality.

\vspace{-2mm}\subsection{Communication overhead}\vspace{-1mm}
We now establish the bit cost model that quantifies the communication overhead 
for each transmission decision. Let $b_{\text{id}}(M_t) = \lceil \log_2 M_t \rceil$ 
be the index cost that captures the number of bits required to reference one of $M_t$ 
prototypes in the shared memory. Let $b_{\text{new}}=8d$ (under $8$ bit quantization) be the cost to 
transmit a new $d$-dimensional latent representation.  The semantic bits transmitted for frame $t$ are:
\vspace{-2mm}\begin{equation}
b_t = H_t \cdot b_{\text{id}}(M_{t-1}) + (1-H_t) \cdot b_{\text{new}} + b_{\text{res},t},
\label{eq:bits}
\vspace{-2mm}\end{equation}
where the first term accounts for index transmission on a hit ($H_t = 1$), 
the second term accounts for new latent transmission on a miss ($H_t = 0$), 
and $b_{\text{res},t} \geq 0$ represents optional residual bits used to encode 
the error $\bm{\epsilon}_t$ when quality enhancement is needed to meet a 
distortion target. In a typical operation, $b_{\text{res},t} = 0$ as the 
prototype or quantized latent alone suffices for reconstruction within the 
target quality. The total SC cost over the video sequence is 
$B_{\text{sem}} = \sum\limits_{t=1}^T b_t$, and the average bits per frame is 
$\bar{B}_{\text{sem}} = B_{\text{sem}}/T$. The hit rate 
$\bar{H} = \frac{1}{T}\sum\limits_{t=1}^T H_t$ measures the fraction of frames that 
successfully reuse stored prototypes. For comparison, a conventional JSCC method would require $B_{\text{raw}}$ bits to transmit the same video 
at equivalent quality, defining the compression ratio 
$\gamma = B_{\text{raw}}/B_{\text{sem}}$.

For MHN-driven SC, our objective is to jointly optimize encoder $f_\theta$ and decoder $g_\phi$ to produce semantically meaningful latent representations enabling effective context reuse while maintaining reconstruction quality and robustness to transmission errors. The optimal threshold $\tau^*$ must balance the rate-distortion-reuse tradeoff, where higher $\tau$ reduces memory hits (higher rate) but ensures better matches (lower distortion), while lower $\tau$ increases reuse (lower rate) but accepts poorer matches. We now formulate the optimization problem.

\vspace{-1mm}\section{Proposed Associative Memory Based SC}\label{sec:memories}
\vspace{-1mm}We propose  MHNs as the shared associative memory because their soft retrieval mechanism allows stable matches near threshold $\tau$, provides robustness through weighted prototype combinations, and minimizes storage by preventing duplicates.
\vspace{-5mm}\subsection{Modern Hopfield network memory}\vspace{-1mm}
For a given query, i.e.,  the latent vector 
$\bm{z}_t$ and stored prototypes $\{\bm{\mu}_j\}_{j=1}^{M_t}$, the Hopfield 
memory computes semantic similarity-weighted activations:
$
w_j = \frac{\exp(\beta \langle \bm{z}_t, \bm{\mu}_j \rangle)}{\sum_{k=1}^{M_t} \exp(\beta \langle \bm{z}_t, \bm{\mu}_k \rangle)},$
and returns a normalized latent vector:
\vspace{-1.5mm}\begin{equation}
\tilde{\bm{z}}_t = \frac{\sum_{j=1}^{M_t} w_j \bm{\mu}_j}{\left\|\sum_{j=1}^{M_t} w_j \bm{\mu}_j\right\|_2},
\label{eq:hopfield_retrieval}
\vspace{-1mm}\end{equation}
where $\beta > 0$ controls retrieval sharpness. As $\beta \to \infty$, the weights concentrate on a single prototype ($w_{j^*} \to 1$), recovering hard nearest-neighbor selection, while lower $\beta$ produces smooth blending across distinct semantic concepts.

Next, we introduce a \emph{reasoning capacity metric} that measures semantic efficiency, i.e. how effectively a context aware SC converts transmitted bits into useful semantic information.

\vspace{-2mm}\subsection{Reasoning capacity metric}
\vspace{-1mm}
We define the \emph{semantic efficiency} as $
\eta= \underbrace{\frac{Q}{T}}_{\text{hit rate}} \times \underbrace{\frac{B_{\text{raw}}}{B_{\text{sem}}}}_{\text{compression ratio}},$
where $H_t \in \{0,1\}$ is the hit indicator at time $t$, 
$Q \triangleq \sum_{t=1}^{T} H_t$ is the total number of hits over a sequence of length $T$, 
$B_{\text{raw}}$ is the number of bits required by a non-semantic (raw) codec, 
and $B_{\text{sem}}$ is the number of bits transmitted by the semantic system.
This metric quantifies the system's ability to reuse semantic content ($Q/T$) while achieving bandwidth savings ($B_{\text{raw}}/B_{\text{sem}}$). Following \cite{chaccour2024less}, the 
corresponding \emph{reasoning capacity} can be written as:
\vspace{-2mm}\begin{equation}
C_R = \Omega \log_2(1+\eta),
\label{eq:CR}
\vspace{-2mm}\end{equation}
where $\Omega > 0$ is a scaling constant (typically $\Omega = 1$ for 
dimensionless capacity, or chosen to map to physical units like bits/s/Hz). 
The logarithmic form ensures: (i)~$C_R$ grows approximately linearly for 
small $\eta$ ($C_R \approx \Omega \eta/\ln 2$ when $\eta \ll 1$), capturing 
incremental improvements; and (ii)~$C_R$ saturates for large $\eta$, avoiding 
unbounded values as compression approaches the theoretical limit. A high reasoning capacity implies that the system 
successfully retrieves stored semantics (high $Q/T$) while minimizing 
transmission overhead (high $B_{\text{raw}}/B_{\text{sem}}$). Memory 
architectures that reduce unnecessary New transmissions near threshold $\tau$ 
increase $Q/T$, directly improving $C_R$.
We define $D(\hat{\bm{x}}_t,\bm{x}_t)$ as a distortion metric (e.g., MSE), with 
$\bar{D} = \frac{1}{T}\sum_{t=1}^T D(\hat{\bm{x}}_t,\bm{x}_t)$ being the average distortion. The SC design 
objective is to maximize reasoning capacity subject to a reconstruction quality constraint. \vspace{-2mm}\subsection{Problem formulation}\vspace{-1mm}
We formulate the MHN-driven SC system design as the following constrained optimization problem:
\vspace{-3mm}\begin{subequations}
\begin{align}
&\max_{\theta,\phi,\tau,\,\{b_{\text{res},t}\}_{t=1}^T} \; C_R(\theta,\phi,\tau) \tag{6}\label{eq_P1}
\\
    \text{s.t.}\quad & \bar{D}(\theta,\phi,\pi_\tau) \le D_0, \label{eq_C1}\\
    & H_t = \mathbf{1}\{\langle z_t, \mu_{j}^* \rangle \ge \tau\}, \quad \forall t \in [T],\label{eq_C2} \\
    & z_t = f_\theta(x_t), \quad \|z_t\|_2 = 1, \quad \forall t, \label{eq_C3}\\
    & \hat{x}_t = g_\phi(\hat{z_t}), \quad \forall t, \label{eq_C4}, b_{\text{res},t} \ge 0, \quad \forall t, \\
    & M_t \le M_{\max}, \quad \forall t~.\label{eq_C5}
\vspace{-4mm}\end{align}
\end{subequations}
The objective in \eqref{eq_P1} maximizes the reasoning capacity. \eqref{eq_C1} enforces a target reconstruction quality $D_0$, while \eqref{eq_C2} defines the hit/miss logic via threshold $\tau$. The remaining constraints~\eqref{eq_C3}-\eqref{eq_C5} capture the encoder–decoder–memory dynamics and bit allocations. Note that $\theta$ and $\phi$ affect both the latent space geometry (hit/miss decisions) and reconstruction quality ($D_0$). Jointly optimizing $(\theta,\phi,\tau)$ requires gradient-based training with differentiable relaxations of $H_t$ (e.g., straight-through estimators or Gumbel-softmax).  For tractability, we reformulate the constrained problem as an unconstrained Lagrangian:
\vspace{-2mm}\begin{equation}
    \mathcal{L}(\theta,\phi,\tau;\lambda) = \frac{1}{T} B_{\text{sem}}(\theta,\phi,\pi_{\tau}) + \lambda\,\big(\bar{D}(\theta,\phi,\pi_{\tau}) - D_0\big),
    \label{eq_L}
\end{equation}
where $\lambda > 0$ is a multiplier for the distortion constraint. We then seek to maximize $\mathcal{L}$ over $(\theta,\phi,\tau)$. 
\vspace{-2mm}\subsection{Solution approach}\vspace{-1mm}
We solve~\eqref{eq_L} using alternating gradient-based optimization:
In step 1, we replace the hard indicator 
$H_t = \mathbbm{1}\{\langle \bm{z}_t, \bm{\mu}_{j^*} \rangle \geq \tau\}$ 
with a smooth approximation $
    \tilde{H}_t = \sigma\big(\alpha (\langle \bm{z}_t, \bm{\mu}_{j^*} \rangle - \tau)\big),$
where $\sigma(\cdot)$ is the sigmoid function and $\alpha > 0$ controls 
sharpness (higher $\alpha$ approaches the hard threshold). This enables 
backpropagation through the hit/miss decision.In step 2, for fixed $\lambda$, encoder-decoder 
parameters are updated via:
\vspace{-2mm}\begin{equation}
    \theta \leftarrow \theta - \eta_\theta \nabla_\theta \mathcal{L}(\theta,\phi,\tau;\lambda), \quad
    \phi \leftarrow \phi - \eta_\phi \nabla_\phi \mathcal{L}(\theta,\phi,\tau;\lambda),\nonumber
\end{equation}
where $\eta_\theta, \eta_\phi > 0$ are learning rates. The threshold $\tau$ 
is updated similarly or via grid search over $[0,1]$ for each $(\theta,\phi)$. Finally,  $\lambda$ is adjusted to enforce the constraint:
\vspace{-2mm}\begin{equation}
    \lambda \leftarrow \max\big(0, \lambda + \eta_\lambda (\bar{D}(\theta,\phi,\tau) - D_0)\big),
\vspace{-2mm}\end{equation}
where $\eta_\lambda > 0$ is the step size. If $\bar{D} > D_0$ (quality 
violation), $\lambda$ increases to penalize distortion more heavily. This procedure alternates between minimizing $\mathcal{L}$ over 
$(\theta,\phi,\tau)$ and updating $\lambda$ until convergence to a saddle 
point satisfying the quality constraint. The details of the solution is outlined in Algorithm~\ref{alg:training}. 

\subsection{Theoretical analysis}
\vspace{-1mm}\begin{theorem}[Rate–Distortion–Reuse Tradeoff]
\label{thm:rdr_tradeoff}
For fixed encoder–decoder $(\theta,\phi)$, the triple $(C_R, \bar{D}, \tau)$ lies on a Pareto frontier 
 $ \mathcal{F} = \{(C_R(\theta,\pi,\tau),\,\bar{D}(\theta,\pi,\tau),\,\tau) : \tau \in (0,1)\},  
$
characterized by:
\begin{enumerate}
    \item \emph{Monotonic relationships:}
    $
    \frac{\partial C_R}{\partial \tau} \!<\! 0, 
    \frac{\partial \bar{D}}{\partial \tau} \!< \!0, 
    \frac{\partial \bar{H}}{\partial \tau} \!<\! 0~.
$
    \item \emph{Optimal threshold:} For a target distortion $D_0$, the optimal reuse threshold is   \vspace{-2mm}\begin{equation}
        \tau^*(D_0) = \inf\{\tau \in (0,1) : \bar{D}(\tau) \le D_0\},
\vspace{-1mm}\end{equation}
    and $\tau^*(D_0)$ is non-increasing in $D_0$.
   \item \emph{Pareto optimality:} If we consider two operating points $(\tau_1, D_1, C_{R,1})$ and $(\tau_2, D_2, C_{R,2})$ on $\mathcal{F}$, then it is not possible for one to dominate the other in both distortion and capacity. Specifically, there is no $(\tau_1, D_1, C_{R,1})$ and $(\tau_2, D_2, C_{R,2})$ on $\mathcal{F}$ such that $D_1 \le D_2$ and $C_{R,1} \ge C_{R,2}$ with at least one strict inequality.
    \end{enumerate}
\end{theorem}
\vspace{-1mm}\begin{IEEEproof}
The proof is omitted due to space constraints.
\end{IEEEproof}
Theorem~\ref{thm:rdr_tradeoff} establishes that all memory 
architectures face the same fundamental constraints. However, the 
\emph{shape and position} of the Pareto frontier depends critically 
on the retrieval mechanism. We now formalize when MHN's 
associative recall provides a quantifiable advantage over VQ's hard 
quantization.
\vspace{-5mm}\begin{theorem}\label{theorem_2}[MHN Bit Savings]
Consider a video sequence where latents exhibit bounded semantic drift around a stored prototype $
    z_t = \mu^* + \delta_t, \text{where } \|\delta_t\|_2 \le \varepsilon \text{ for some } \mu^* \in \mathcal{M} \text{ and small } \varepsilon > 0.
$
Assume all prototypes are unit-norm and lie on the sphere $\mathbb{S}^{d-1}$. For a given threshold $\tau \in (0,1)$ and Hopfield inverse temperature $\beta > 0$, let $N_{\text{refresh}}^H$ and $N_{\text{refresh}}^V$ denote the number of memory refresh events (new transmissions) for Hopfield and VQ respectively over a horizon of $T$ frames. Under the following conditions:
\begin{enumerate}
    \item The prototypes $\{\mu_j\}_{j=1}^M$ satisfy minimum separation: $\min_{j \neq j'} \|\mu_j - \mu_{j'}\|_2 \ge \Delta > 0$.
    \item The drift $\delta_t$ is i.i.d. with $\mathbb{E}[\delta_t] \!=\! 0$ and $\mathbb{E}[\|\delta_t\|_2^2] \!= \!\sigma^2 \le \varepsilon^2$.
    \item The threshold satisfies $\tau < 1 - \varepsilon - \frac{\varepsilon^2}{2}$ (boundary regime).
\end{enumerate}
Then, number of memory updates for MHN-driven SC is lower than that of VQ-VAE and is bounded as:
\vspace{-2mm}\begin{equation}
\mathbb{E}[N_{\text{refresh}}^H] \le \mathbb{E}[N_{\text{refresh}}^V] \cdot \left( 1 - \frac{\beta \varepsilon^2}{2d} \right) + O(\varepsilon^3)
\vspace{-2mm}\end{equation}
where the expectation is over the drift realizations $\{\delta_t\}_{t=1}^T$.
\end{theorem}
\vspace{-1mm}\begin{IEEEproof}
See Appendix~\ref{proof_theorem_2}.    
\end{IEEEproof}
\vspace{-2mm}\begin{corollary}\label{corollary}
Under the conditions of Theorem~\ref{theorem_2}, the expected semantic bit savings are:
\vspace{-4mm}\begin{equation}
\mathbb{E}[B_{\text{sem}}^V - B_{\text{sem}}^H] \geq (b_{\text{new}} - \overline{\log_2 M}) \cdot \frac{\beta\epsilon^2 T}{2d} \mathbb{E}[N_{\text{refresh}}^V].
\end{equation}
\vspace{-5mm}\end{corollary}
\vspace{-3mm}\begin{IEEEproof}
    See Appendix~\ref{corollary_proof}.
\end{IEEEproof}

\setlength{\textfloatsep}{0pt}\begin{algorithm}[t]
\caption{\small Training Procedure for Memory-Based SC}
\label{alg:training}
\vspace{-1.5mm}\begin{algorithmic}[1]\scriptsize
\REQUIRE Training video data $\{\bm{x}_t\}_{t=1}^T$, target distortion $D_0$
\ENSURE Encoder $f_\theta$, decoder $g_\phi$, threshold $\tau$
\STATE Initialize $\theta, \phi$ randomly, $\tau = 0.7$, $\lambda = 1.0$
\STATE Initialize memory $\mathcal{M}_0 = \emptyset$

\FOR{epoch $= 1$ to $N_{\text{epochs}}$}
    \FOR{batch $\{\bm{x}_t\}_{t=1}^B$}
        \STATE Encode: $\bm{z}_t = f_\theta(\bm{x}_t) / \|f_\theta(\bm{x}_t)\|_2$
        \STATE Compute hits: $\tilde{H}_t = \sigma(\alpha(\langle \bm{z}_t,\bm{\mu}_{j^*}\rangle - \tau))$
        \STATE Retrieve/transmit: $\hat{\bm{z}}_t$ via Eq.~\eqref{eq:decoder_input}
        \STATE Decode: $\hat{\bm{x}}_t = g_\phi(\hat{\bm{z}}_t)$
        \STATE Compute $\mathcal{L}(\theta,\phi,\tau;\lambda)$ via Eq.~\eqref{eq_L}
        \STATE Update: $\theta \leftarrow \theta - \eta_\theta \nabla_\theta \mathcal{L}$
        \STATE Update: $\phi \leftarrow \phi - \eta_\phi \nabla_\phi \mathcal{L}$
        \STATE Update: $\tau \leftarrow \tau - \eta_\tau \nabla_\tau \mathcal{L}$ (or grid search)
        \STATE Update memory: $\mathcal{M} \leftarrow \mathcal{M} \cup \{\bm{z}_t : H_t = 0\}$
    \ENDFOR
    \STATE Update multiplier: $\lambda \leftarrow \max(0, \lambda + \eta_\lambda(\bar{D} - D_0))$
\ENDFOR
\RETURN $f_\theta, g_\phi, \tau$
\vspace{-1mm}\end{algorithmic}
\end{algorithm}

\vspace{-2mm}\section{Simulation Results and Analysis}\vspace{-2mm}
We evaluate the proposed SC system on a variety of video content types and compare it against a VQ-VAE based SC \cite{fu2023vqsc, 
miao2024vqdeepvsc} baseline. We implement a lightweight learned encoder–decoder (e.g., a $\beta$-VAE \cite{VAE} or a shallow CNN autoencoder) that produces $d$-dimensional latent vectors $z_t$, normalized to unit norm. Unless otherwise specified, new latent vectors are quantized with $8$ bits per value (i.e., $b_{\text{new}} = 8d$). All comparisons between Hopfield and VQ memories are conducted under matched reconstruction fidelity: we calibrate the similarity threshold $\tau$ for each method such that the peak signal-to-noise ratio (PSNR) of the reconstructed video falls within a $\pm0.2$\,dB range for both methods.

\vspace{-2mm}\subsection{Dataset description}\vspace{-2mm}
We use synthetic and real video clips from the imageio library~\cite{imageio-zenodo} across eight content regimes: \emph{static}, \emph{low\_motion}, \emph{high\_motion}, \emph{scene\_change}, \emph{repetitive\_return}, \emph{cam\_shake}, \emph{high\_texture}, and \emph{illum\_change}, with $T \in [36, 80]$ frames. We test three scenarios: (A) \emph{STABLE}, where frames remain highly similar  over time (\emph{static}, \emph{low\_motion}, \emph{repetitive\_return}); (B) \emph{GRADUAL DRIFT}, where changes accumulate slowly from lighting, camera motion, or appearance variations (\emph{illum\_change}, minor \emph{cam\_shake}); and (C) \emph{MODERATE}, with intermediate dynamics between stable and drifting content (\emph{high\_motion}, \emph{high\_texture}).
We evaluate performance by varying threshold $\tau$, memory size $M_{\max}$, 
and robustness to memory corruption. 

\vspace{-2mm}\subsection{Bandwidth savings}\label{sec:results}\vspace{-2mm}
\begin{figure}[t]
 \vspace{-3mm}   \centering
 \begin{subfigure}[b]{0.23\textwidth}
\includegraphics[width=\textwidth,height=3.75cm]{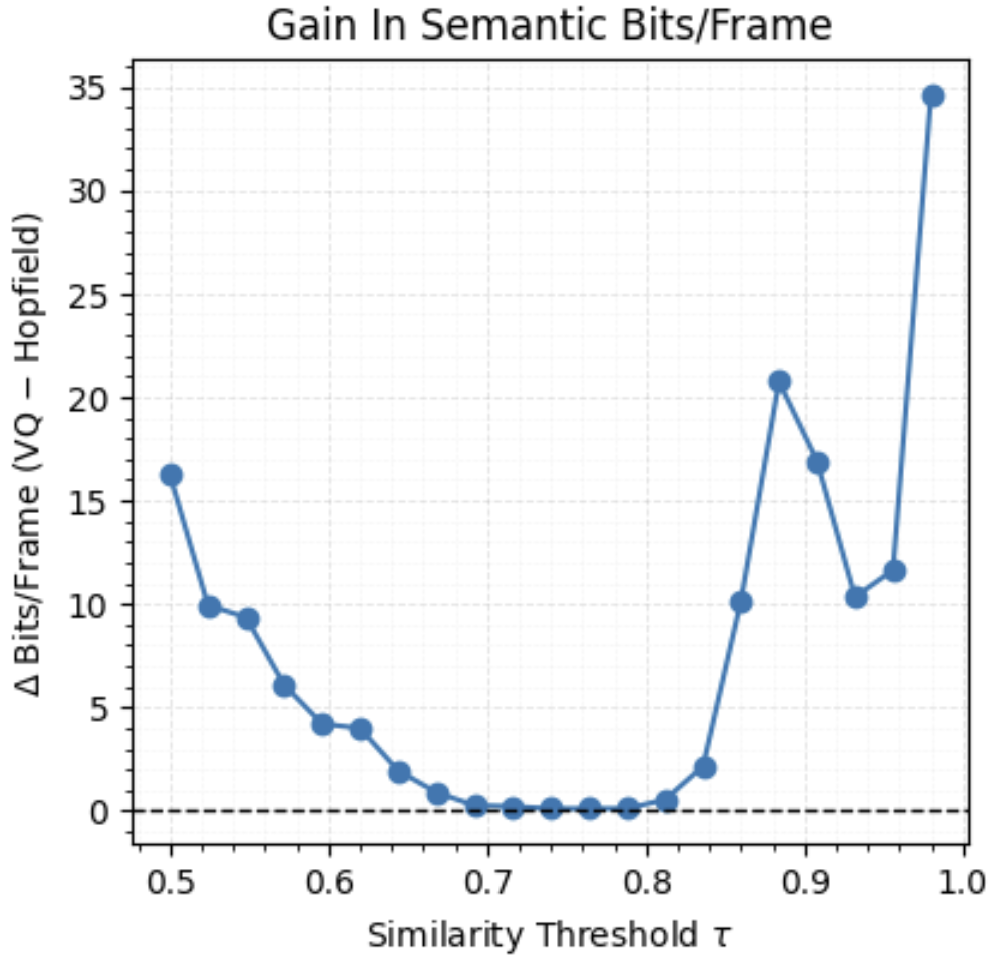}
\vspace{-6mm}\caption{}
    \label{fig:bits_gain}
    \vspace{-3mm}
    \end{subfigure}
    \begin{subfigure}[b]{0.25\textwidth}
    \includegraphics[width=\textwidth,height=3.75cm]{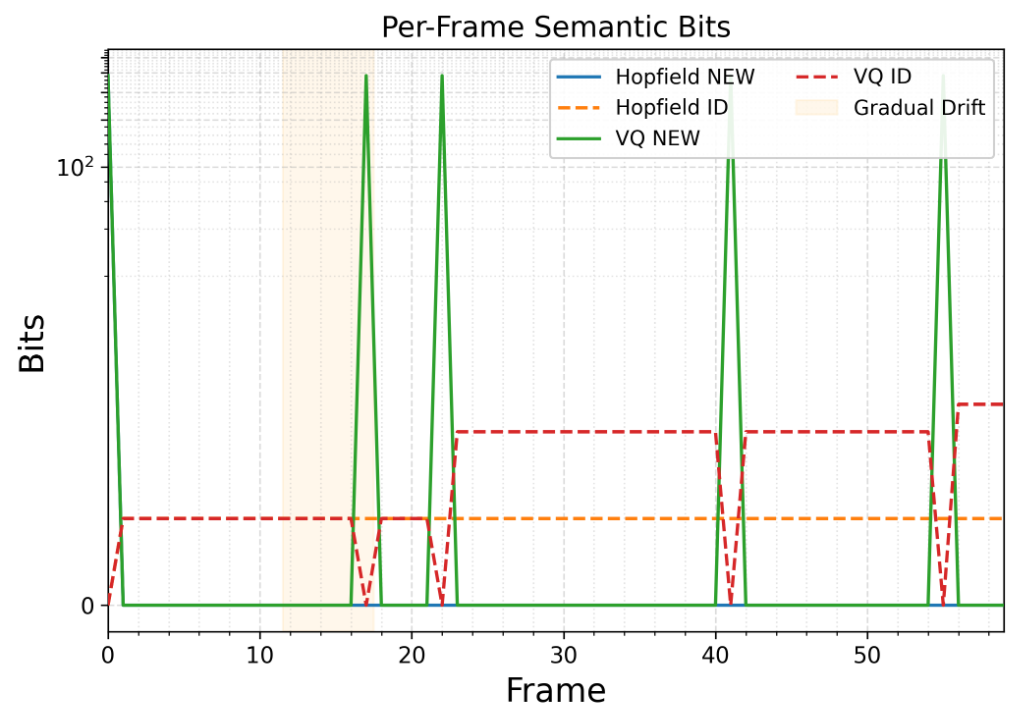}
\vspace{-6mm}\caption{ }
    \label{fig:id_gain}
    \vspace{-3mm}
    \end{subfigure}
    \caption{\small (a) Amount of reduction in bits transmitted. (b) Illustration of new latent vector transmissions for VQ-VAE compared to stable knowledge base for MHN.}\vspace{-2mm}
\end{figure}

\begin{table}[t]
  \centering
  \caption{\small Bits/frame for diverse content types.}
  \label{tab:byproto}
 \vspace{-2mm} \setlength{\tabcolsep}{3pt}
  \footnotesize
\resizebox{0.9\columnwidth}{!}{%
  \begin{tabular}{ll
                  S[table-format=3.2(2)]
                  S[table-format=1.3(2)]
                  S[table-format=2.2(2)]}
    \toprule
    Protocol & Method & {Bits/frame} & {Hit rate} & {$C_R$} \\
    \midrule
    \multirow{2}{*}{STABLE}
      & Hopfield & 9.51 & 0.978  & 14.26  \\
      & VQ       & 9.51 & 0.978  & 14.26  \\
    \midrule
    \multirow{2}{*}{GRADUAL DRIFT}
      & Hopfield & 68.66  & 0.825  & 14.26 \\
      & VQ       & 228.53  & 0.408  & 9.04 \\
    \midrule
    \multirow{2}{*}{MODERATE}
      & Hopfield & 9.51  & 0.978  & 14.26  \\
      & VQ       & 9.51  & 0.978  & 14.26  \\
    \bottomrule
  \end{tabular}%
  }
  \label{tab:CR}\vspace{-3mm}
\end{table}
Table~\ref{tab:byproto} shows the bits per frame and reasoning capacity for 
 three scenarios. In the \emph{STABLE} and \emph{MODERATE} scenarios, MHN and VQ\mbox{-}VAE perform equivalently, with 9.51 bits/frame, hit rate \(0.978\), and \(C_R=14.26\). In the \emph{GRADUAL DRIFT} scenario (slow concept shift), MHN transmits about \(70\%\) fewer bits per frame than VQ\mbox{-}VAE (\(68.66\) vs.\ \(228.53\)) while maintaining a higher hit rate (\(0.825\) vs.\ \(0.408\)) and stronger compression (\(C_R=14.26\) vs.\ \(9.04\)). Fig.~\ref{fig:bits_gain} shows the gain in the average amount of bits communicated for MHN-driven SC compared to VQ-VAE baseline for varying similarity threshold $\tau$. Clearly, our MHN approach outperforms VQ-VAE when there is a gradual change in semantics across the frames. Fig.~\ref{fig:id_gain} shows that VQ-VAE baseline requires frequent updates to the knowledge base, which are illustrated by the spikes corresponding to new latent vector transmissions (see legends "VQ NEW" vs "Hopfield NEW").

In scenario A, Hopfield exhibits a much lower frequency of new transmissions than VQ for the same quality, yet the index entropy $H_S$ of Hopfield remains comparable to that of VQ. This indicates that Hopfield’s bit savings are not coming from a trivial solution like using one index for everything (which would collapse $H_S$); instead, Hopfield effectively recalls a variety of prototypes as needed but avoids creating extraneous new ones. In contrast, in highly non-stationary or texture rich segments (\emph{high\_motion}, \emph{high\_texture}), both Hopfield and VQ end up sending new content frequently (low reuse), and thus their performance converges (with high $H_S$ for both, reflecting many distinct indices used).

\vspace{-2mm}\subsection{Reuse vs.\ diversity tradeoff}\vspace{-2mm}
\begin{figure}[t]
    \centering
\includegraphics[width=0.35\textwidth,height=4.5cm]{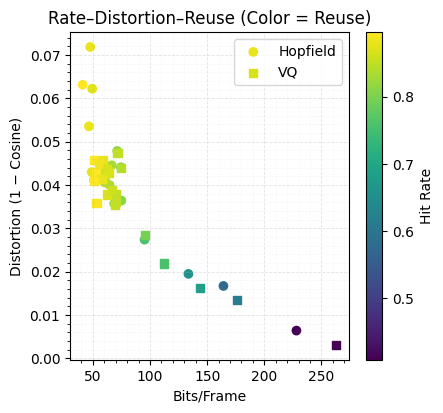}
\vspace{-3mm}\caption{\small Rate-distortion-reuse tradeoff.}
    \label{fig:rdr}
    \vspace{-2mm}
\end{figure}
Fig.~\ref{fig:rdr} shows that Hopfield networks consistently achieve lower transmission rates (bits/frame)than VQ at comparable distortion levels; the advantage increases as the hit rate decreases, reaching up to \(\sim\!15\text{--}25\%\) bit savings in the low-hit-rate regime. (dark blue points). The soft associative retrieval in Hopfield allows better interpolation between stored prototypes, reducing the need to transmit new representations compared to VQ's hard selection. This demonstrates that Hopfield's continuous retrieval mechanism provides a tangible rate-distortion advantage for SC in video streaming scenarios.
\vspace{-2mm}\subsection{Robustness to Rx memory errors}
\vspace{-2mm}
\begin{figure}[t]
    \centering
    \begin{subfigure}[b]{0.24\textwidth}
        \centering
        \includegraphics[width=\textwidth,height=3.75cm]{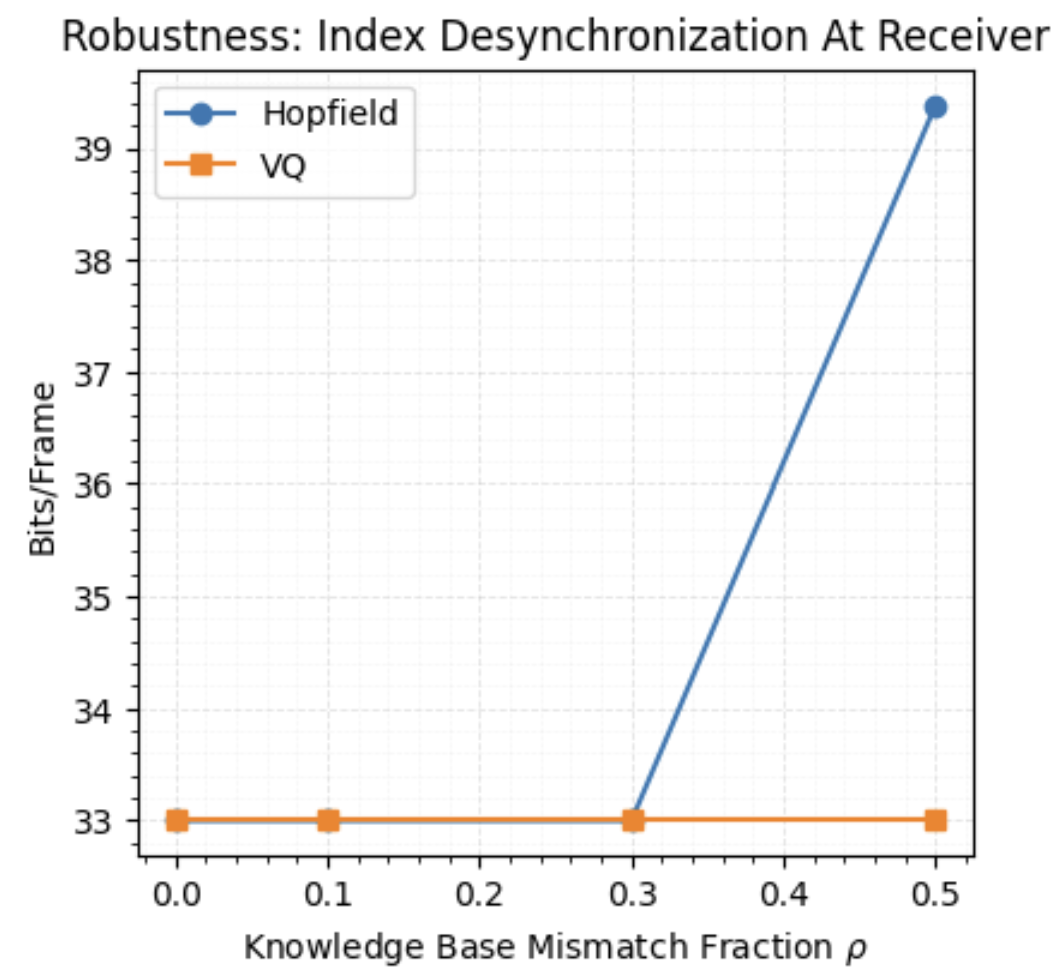}
       \vspace{-6mm} \caption{}
        \label{fig:id_desync}
    \end{subfigure}
    \hfill
    \begin{subfigure}[b]{0.24\textwidth}
        \centering
        \includegraphics[width=\textwidth,height=3.65cm]{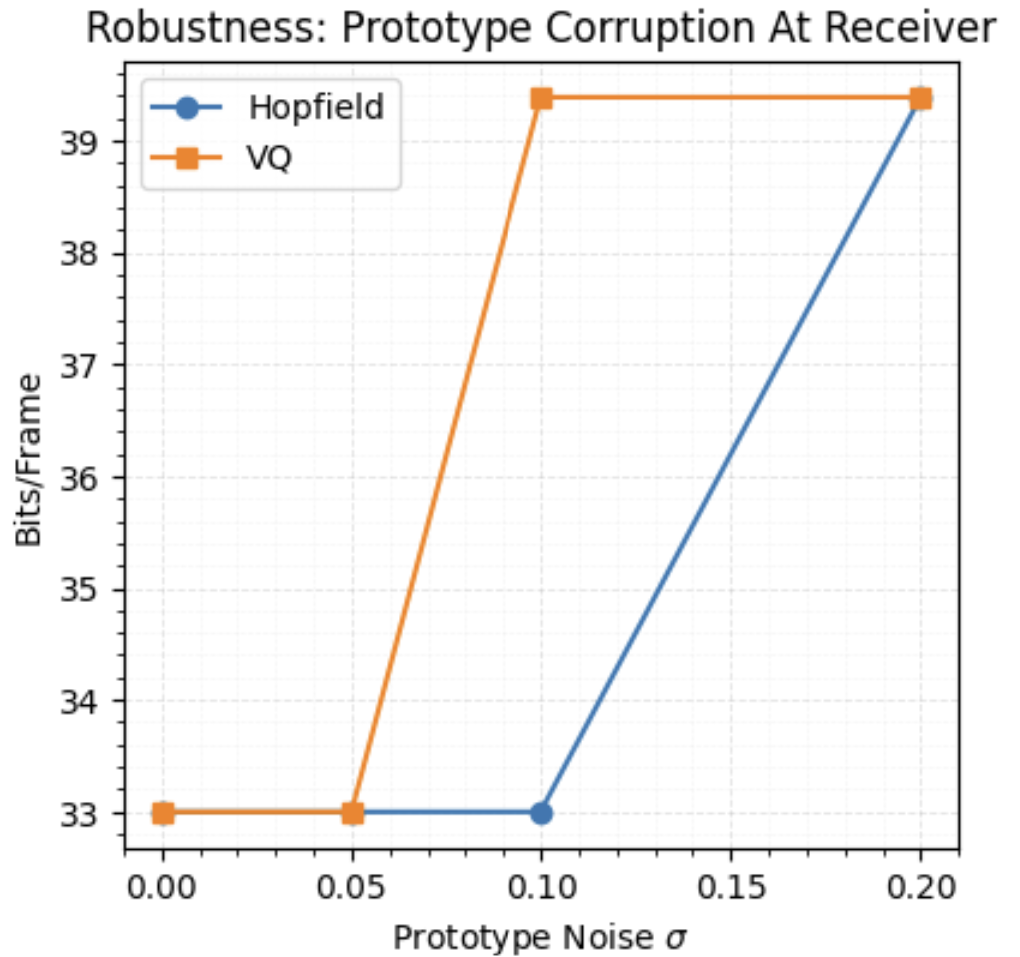}
    \vspace{-6mm}\caption{}
    \label{fig:vec_perturb}
    \end{subfigure}
    \vspace{-7mm}\caption{\small (a) Robustness of MHN-driven SC to knowledge base retrieval mechanism errors, and (b) stored prototype mismatches.   }
    \label{fig:combined}
\vspace{-0mm}\end{figure}
We evaluate robustness under two Rx memory corruption modes: (a) ID-map de-synchronization, where the mapping from semantic IDs to memory slots is randomly permuted for fraction $p$ of entries, and (b) vector perturbation, where stored prototypes are corrupted by additive Gaussian noise $\epsilon_j \sim \mathcal{N}(0, \sigma^2 \mathbf{I})$ and re-normalized. The Rx verifies each ID transmission using a short cue (SimHash fingerprint of $B_{\text{cue}} = 32$ bits, tolerating up to $25\%$ bit errors) and performs content-addressable lookup on cue f
, requesting a refresh if verification still fails. Figures~\ref{fig:id_desync} and~\ref{fig:vec_perturb} show that Hopfield's associative retrieval offers a semantic robustness advantage in the regime of mild memory corruption ($p \leq 0.1$, $\sigma \leq 0.05$), incurring fewer refreshes than VQ's hard nearest-centroid lookup and thus saving bits for example, achieving $\Delta\text{bits} = +381$ at $p = 0$ versus $-381$ at $p = 0.5$, demonstrating that soft attractor dynamics tolerate small misalignments better than hard quantization near decision boundaries. However, under severe corruption ($\sigma \geq 0.10$), the Hopfield attractor can be misled by averaging corrupted neighbors, causing the advantage to narrow or reverse, highlighting a tradeoff between semantic tolerance and robustness that can be tuned via cue length, association sharpness $\beta$, and similarity threshold $\tau$.

\vspace{-1mm}\section{Conclusion}\vspace{0mm}

In this paper, we have presented a memory-augmented SC framework leveraging MHNs as shared associative memories between Tx and Rx. The proposed soft attention-based retrieval mechanism maintains stable matching decisions under gradual semantic drift, avoiding the bandwidth waste caused by baseline methods' hard quantization approach. We introduced a reasoning capacity metric that quantifies semantic efficiency as the product of memory reuse rate and compression ratio. Theoretical analysis established the fundamental rate-distortion-reuse tradeoff and proved that soft retrieval reduces unnecessary transmissions compared to hard quantization under bounded semantic drift conditions. Extensive simulations demonstrated that the MHN-based approach achieves approximately $14\%$ average bit reduction across diverse video scenarios and up to $70\%$ savings under gradual content changes, while maintaining  reconstruction quality and exhibiting graceful degradation under transmission errors. 

\vspace{-1mm}\bibliographystyle{IEEEbib}
\def\baselinestretch{0.89}
\vspace{-2mm}
\bibliography{refs}
\appendices

\vspace{-3mm}\appendices 
\section{Proof of Theorem 2}
\label{proof_theorem_2}
\vspace{-2mm}
The expected refresh counts are
$
\mathbb{E}[N_{\text{refresh}}^V] = T \cdot P^V_{\text{miss}}, \quad \mathbb{E}[N_{\text{refresh}}^H] = T \cdot P^H_{\text{miss}}
$,
where $P^V_{\text{miss}} = \mathbb{P}(\text{Miss}_t^V = 1)$ and $P^H_{\text{miss}} = \mathbb{P}(\text{Miss}_t^H = 1)$. Given $z_t = \mu^* + \delta_t$, the similarity to the correct prototype is:
$
s^*_t = \langle z_t, \mu^* \rangle = \langle \mu^* + \delta_t, \mu^* \rangle = \langle \mu^*, \mu^* \rangle + \langle \delta_t, \mu^* \rangle = 1 + \langle \delta_t, \mu^* \rangle
$
(using $\|\mu^*\|_2 = 1$).
For $j \neq j^*$, we define $\rho = \max_{j \neq j^*} \langle \mu^*, \mu_j \rangle$. By the minimum separation assumption, we have $
\|\mu_j - \mu^*\|_2^2 = 2 - 2\langle \mu_j, \mu^* \rangle \geq \Delta^2.
$
Therefore,
$
\langle \mu_j, \mu^* \rangle \leq 1 - \frac{\Delta^2}{2} =: \rho < 1
$.
Therefore, the similarity to a competing prototype is
\vspace{-2mm}\begin{equation}
    \langle z_t, \mu_j \rangle = \langle \mu^*, \mu_j \rangle + \langle \delta_t, \mu_j \rangle \leq \rho + \epsilon
\vspace{-2mm}\end{equation}
This follows from Cauchy-Schwarz inequality $|\langle \delta_t, \mu_j \rangle| \leq \|\delta_t\|_2 \|\mu_j\|_2 \leq \epsilon$. Under the assumption that $\mu^*$ is the closest prototype (satisfied when $\epsilon$ is small relative to prototype separation), the miss condition simplifies to:
$
1 + \langle \delta_t, \mu^* \rangle < \tau
$.
Therefore, we obtain
$
P^V_{\text{miss}} = \mathbb{P}\left(\langle \delta_t, \mu^* \rangle < \tau - 1\right)
$.
We now derive the {associative boost} that MHN  provides through its soft attention mechanism.
\subsubsection{ Compute soft attention weights}

The attention weight for the correct prototype is:
\vspace{-2mm}\begin{equation}
    w_{j^*}(\beta) = \frac{\exp\left(\beta(1 + \langle \delta_t, \mu^* \rangle)\right)}{\exp\left(\beta(1 + \langle \delta_t, \mu^* \rangle)\right) + \sum\limits_{j \neq j^*} \exp\left(\beta \langle z_t, \mu_j \rangle\right)}.
    \label{eq_step1}
\vspace{-1mm}\end{equation}
For simplicity, let $\xi = \langle \delta_t, \mu^* \rangle$ and $\xi_j = \langle \delta_t, \mu_j \rangle$ for $j \neq j^*$, and
factoring out $\exp(\beta(1 + \xi))$, 
and defining $\gamma_j = \langle \mu^*, \mu_j \rangle - 1 \leq \rho - 1 < 0$ (since prototypes are separated), \eqref{eq_step1} becomes
\vspace{-3mm}\begin{equation}
w_{j^*}(\beta) = \frac{1}{1 + \sum_{j \neq j^*} \exp\left(\beta(\gamma_j + \xi_j - \xi)\right)}.
\vspace{-1mm}\end{equation}
For small $\epsilon$, $|\xi|, |\xi_j| \leq \epsilon$, 
$\gamma_j + \xi_j - \xi \leq (\rho - 1) + 2\epsilon < 0
$
(using the boundary assumption), which implies $1 - \rho > 2\epsilon$. Further, performing a Taylor expansion around $\xi = \xi_j = 0$,
Let $\alpha = \frac{1}{1 + (M-1)\exp(\beta(\rho-1))}$ be the weight when $\delta_t = 0$.
For small $\epsilon$:
$
w_{j^*}(\beta) \approx \alpha \left(1 + \beta\xi + O(\epsilon^2)\right)
$. Following similar steps, we obtain for competing prototypes,
$$
w_j(\beta) \approx \frac{1-\alpha}{M-1} \left(1 - \beta\xi + O(\epsilon^2)\right) \quad \text{for } j \neq j^*
$$

\subsubsection{Similarity after MHN Update}

We write
$
\langle \tilde{z}_t, \mu^* \rangle = \sum_{j=1}^M w_j(\beta) \langle \mu_j, \mu^* \rangle 
= w_{j^*}(\beta) \cdot 1 + \sum_{j \neq j^*} w_j(\beta) \langle \mu_j, \mu^* \rangle
$.
Using the approximations derived above, we can write it as:
\begin{equation}
\begin{aligned}
\langle \tilde{z}_t, \mu^* \rangle
&\approx \alpha(1 + \beta\xi) + (1-\alpha)(1-\beta\xi)\rho \\
&= \alpha(1-\rho) + \rho + \beta\xi\big[\alpha - (1-\alpha)\rho\big]
\end{aligned}
\end{equation}

For large $\beta$ or well-separated prototypes, $\alpha \approx 1$, so:
$
\langle \tilde{z}_t, \mu^* \rangle \approx 1 + \beta\xi + O(\epsilon^2)
$.
Including second-order terms in the Taylor expansion (derived from the exponential in the weights and the normalization):
\vspace{-2mm}\begin{equation}
    \langle \tilde{z}_t, \mu^* \rangle \approx 1 + \langle \delta_t, \mu^* \rangle + \frac{\beta}{2d} \|\delta_t\|_2^2 + O(\epsilon^3)
\vspace{-2mm}\end{equation}
$\frac{\beta}{2d}\|\delta_t\|_2^2$ arises from the curvature of the exponential softmax,  averaging over $d$ dimensions, and is the normalization factor.

Next, we compute the Hopfield miss probability. 
Hopfield misses when
$
\langle \tilde{z}_t, \mu^* \rangle < \tau
$.
Using the boost,
$
1 + \langle \delta_t, \mu^* \rangle + \frac{\beta}{2d}\|\delta_t\|_2^2 < \tau
$.
Rearranging gives
$
\langle \delta_t, \mu^* \rangle < \tau - 1 - \frac{\beta}{2d}\|\delta_t\|_2^2.
$
Since $\|\delta_t\|_2^2 \geq 0$:
\vspace{-2mm}\begin{equation}
P^H_{\text{miss}} = \mathbb{P}\left(\langle \delta_t, \mu^* \rangle < \tau - 1 - \frac{\beta}{2d}\|\delta_t\|_2^2\right).
\label{eq_P_Miss_H}
\vspace{-1mm}\end{equation}
\eqref{eq_P_Miss_H} can be bounded as $
P^H_{\text{miss}} \leq \mathbb{P}\left(\langle \delta_t, \mu^* \rangle < \tau - 1\right) = P^V_{\text{miss}}.$
For tighter analysis, use $\|\delta_t\|_2^2 \leq \epsilon^2$, which leads to
\vspace{-2mm}\begin{equation}
  P^H_{\text{miss}} \leq \mathbb{P}\left(\langle \delta_t, \mu^* \rangle < \tau - 1 - \frac{\beta\epsilon^2}{2d}\right).  
\vspace{-1mm}\end{equation}
\subsubsection{Ratio of miss probabilities}
The VQ miss probability can be written as
$
P^V_{\text{miss}} = \int_{-\infty}^{\tau-1} p(\xi) \, d\xi
$.
For Hopfield (to first order in $\epsilon^2$), this becomes
$
P^H_{\text{miss}} \approx \int_{-\infty}^{\tau-1-\frac{\beta\epsilon^2}{2d}} p(\xi) \, d\xi
$. Further, the difference can be computed as
\vspace{-2mm}\begin{equation}
P^V_{\text{miss}} - P^H_{\text{miss}} \approx \int_{\tau-1-\frac{\beta\epsilon^2}{2d}}^{\tau-1} p(\xi) \, d\xi
\label{eq_diff}
\end{equation}
For small $\frac{\beta\epsilon^2}{2d}$, \eqref{eq_diff} can be approximated as
$
\approx p(\tau-1) \cdot \frac{\beta\epsilon^2}{2d}
$.
The ratio becomes
$
\frac{P^H_{\text{miss}}}{P^V_{\text{miss}}} = 1 - \frac{p(\tau-1)}{P^V_{\text{miss}}} \cdot \frac{\beta\epsilon^2}{2d}.
$
We define $C = \frac{p(\tau-1)}{P^V_{\text{miss}}}$ (a constant depending on the distribution of $\delta_t$ and threshold $\tau$).
Then
$
P^H_{\text{miss}} = P^V_{\text{miss}} \left(1 - C\frac{\beta\epsilon^2}{2d}\right) + O(\epsilon^3)
$. Since the drift $\delta_t$ is i.i.d. across frames, we can write after aggregating over horizon $T$ as:
$
\mathbb{E}[N_{\text{refresh}}^V] = \sum_{t=1}^T \mathbb{E}[\text{Miss}_t^V] = T \cdot P^V_{\text{miss}}
$, and 
$
\mathbb{E}[N_{\text{refresh}}^H] = \sum_{t=1}^T \mathbb{E}[\text{Miss}_t^H] = T \cdot P^H_{\text{miss}}
$.
Substituting the ratio, we get
\vspace{-2mm}\begin{equation}
\begin{aligned}
    \mathbb{E}[N_{\text{refresh}}^H] = \mathbb{E}[N_{\text{refresh}}^V] \left(1 - C\frac{\beta\epsilon^2}{2d}\right) + O(\epsilon^3)
\end{aligned}
\end{equation}
Absorbing the constant $C$ (which depends on the drift distribution but is $O(1)$):
\vspace{-2mm}\begin{equation}
    \mathbb{E}[N_{\text{refresh}}^H] \leq \mathbb{E}[N_{\text{refresh}}^V] \cdot \left(1 - \frac{\beta\epsilon^2}{2d}\right) + O(\epsilon^3)
\vspace{-1mm}\end{equation}



This completes the proof of Theorem~\ref{theorem_2}. \qed

\vspace{-2mm}\section{Proof of Corollary~\ref{corollary}(Bit Savings)}\label{corollary_proof}

\begin{proof}
From ~\eqref{eq:bits}, each refresh (miss) costs $b_{\text{new}}$ bits (to send a new latent), while a hit costs $\overline{\log_2 M}$ bits (to send an index).
The difference in semantic bits between VQ and Hopfield is
$\mathbb{E}[B_{\text{sem}}^V - B_{\text{sem}}^H] = (b_{\text{new}} - \overline{\log_2 M}) \cdot \mathbb{E}[N_{\text{refresh}}^V - N_{\text{refresh}}^H]
$. From Theorem~\ref{theorem_2}, we have
$
\mathbb{E}[N_{\text{refresh}}^V - N_{\text{refresh}}^H] = \mathbb{E}[N_{\text{refresh}}^V] \cdot \frac{\beta\epsilon^2}{2d} + O(\epsilon^3)
$.
Therefore:
\vspace{-2mm}\begin{equation}
\mathbb{E}[B_{\text{sem}}^V - B_{\text{sem}}^H] = (b_{\text{new}} - \overline{\log_2 M}) \cdot \frac{\beta\epsilon^2}{2d} \mathbb{E}[N_{\text{refresh}}^V] + O(\epsilon^3)
\end{equation}
Dropping higher-order terms yields the stated result.
\end{proof}

\end{document}